\documentstyle[aps,preprint]{revtex}
\begin{document}
\draft
\title{The terahertz phonon laser: a full quantum treatment}
\author{I. Camps$^\dag$, S. S. Makler$^{\dag\ddag}$, H. M. Pastawski$^\S$ and L. E.
F. Fo\'{a} Torres$^\S$}
\address{$^{\dag }$ Instituto de F\'{\i}sica, Universidade Federal Fluminense, Niter\'{o}i, RJ, Brazil}
\address{$^{\ddag }$ Instituto de F\'{\i}sica, Universidade Estadual do Rio deJaneiro, RJ, Brazil}
\address{$^{\S }$ FaMAF, Universidad Nacional de C\'{o}rdoba, 5000 C\'{o}rdoba,Argentina}
\date{\today}
\maketitle

\begin{abstract}
The aim of this work is to describe the behavior of a device capable to
generate high frequency ($\sim THz$) acoustic phonons. This device consists
in a {\it GaAs-AlGaAs} double barrier heterostructure that, when an external
bias is applied, produces a high rate of longitudinal optical $LO$ phonons.
These $LO$ phonons are confined and they decay by stimulated emission of a
pair of secondary longitudinal optical ($\widetilde{LO}$) and transversal
acoustic ($TA$) phonons. The last ones form an intense beam of coherent
acoustic phonons. To study this effect, we start from a tight binding
Hamiltonian that take into account the electron-phonon (e-ph) and
phonon-phonon (ph-ph) interactions. We calculate the electronic current
through the double barrier and we obtain a set of five coupled kinetic
equations that describes the electron and phonon populations. The results
obtained here confirm the behavior of the terahertz phonon laser, estimated
by rougher treatments \cite{jpcm}.
\end{abstract}

\section{Introduction}

Since the pioneering experiments of Chang, Esaki and Tsu \cite{Esaki}, the
tunneling processes in semiconductor heterostructures have attracted great
interest. This has been increasing with the development of modern deposition
techniques such as metal-organic chemical-vapor deposition (MOCVD) and
molecular beam epitaxy (MBE) that can provide artificial nanostructures.
Such structures, coupled to electrical leads, can display many unusual
behaviors.

Most of works have focalized on electron transport phenomena through
heterostructure devices. The double barrier heterostructure (DBH) permits
the creation of several kinds of electronic devices such as semiconductor
laser diodes, ultra-high-frequency oscillators, and many others \cite
{Capasso}.

One of the most important problems in the study of resonant tunneling in DBH
is the e-ph interaction. After the works of Goldman, Tsui and Cunningham 
\cite{Gold1,Gold2}, it was realized the importance of electron-phonon
interaction on the electronic properties of these structures. However,
little importance has been given to the study of the phonons generated in
this process, the way they propagate, their decay processes, etc.

Our purpose is to study the phonon generation in a DBH under the effect of
an external applied bias. Our device consists in a DBH made of {\it %
GaAs-AlGaAs}. It was designed in such a way that the difference $\Delta
\varepsilon $ between the first excited level $\varepsilon _{1}$ and the
ground state $\varepsilon _{0}$ in the well is of the order of the $LO$
phonon energy $\hbar \omega _{1}$, for a small applied bias $V$. Here, $%
\omega _{1}$ is the $\Gamma $-point $LO$ phonon frequency. The Fermi level $%
\varepsilon _{F}^{L}$ in the emitter is such that the excited level is above
it. With a further increase in $V$, the levels in the well are lowered
relative to $\varepsilon _{F}^{L}$. When the ground state comes under the
bottom of the conduction band, the current is almost suppressed until the
level $\varepsilon _{1}$ reaches $\varepsilon _{F}^{L}$. To continue
increasing $V$, the current begins to flow through the excited level, but
since $\Delta \varepsilon $ remains less than $\hbar \omega _{1}$, the
phonon emission is inhibited. For a given bias $V$ the resonant condition $%
\Delta \varepsilon \approx \hbar \omega _{1}$ is achieved and the electrons
begin to decay to the ground state by emitting primary$\ LO$ phonons. The
potential profile and the level positions at this resonant condition are
shown in figure \ref{profile-Fig1}.

For an $Al$ concentration greater than $0.25$ \cite{Juss} or $0.3$ \cite
{Jacob}, these phonons are confined inside the well (they can be also
absorbed by exciting electrons from $\varepsilon _{0}$ to $\varepsilon _{1}$%
).

The process described above, acts in parallel with the decay of primary $LO$
phonons due to anharmonicity. One product of this decay is a secondary
longitudinal optical phonon ($\widetilde{LO}$), the other is a transversal
acoustic ($TA$) phonon \cite{Vall2,Vall}. The $\widetilde{LO}-TA$ pair is
produced by stimulated emission. Therefore these $TA$ phonons, for a bias
greater than the operation threshold, are coherent and form a beam that we
call saser by analogy with a laser. The coherence of the device presented
here was studied previously \cite{ssc1} by developing a formalism similar to
that employed usually to study the coherence in lasers \cite{Haken-p1,Haken2}
which take into account the competition among the several emitted modes.

In the early 60's \cite{Tang}, generation of coherent phonons was discussed
theoretically and observed experimentally \cite{Tucker,Giord}. Several
experiments and applications were suggested \cite{Garmire}. These phonons
can be generated by intense pulses of laser \cite
{Kurz1,Cheng,Kurz2,Kurz3,Kurz4}. More details were described in the review
of Kurz \cite{Kurz5}. In a recent review, Merlin \cite{Merlin} discusses the
generation of pulses of coherent $THz$ optical phonons by intense
femtosecond-pulsed lasers.

Coherent acoustic phonons can be also obtained with a wavelength given by
the period of the superlattice where phonons are optically generated (i.e., $%
500$ {\it \AA }) \cite{Sup}. In our case, the system generates coherent
sound by electronic means with much shorter wavelength. The generation of
high-frequency mono\-chromatic acoustic waves by laser-induced
thermomodulation was reported by Damen \cite{Damen}. The phonon frequency
ranged from $2$ to $4\ GHz$ comparing with $2\ THz$ for our device.

Generation of phonons by stimulated emission was theoretically considered 
\cite{Haken1} and observed experimentally. Bron and Grill \cite{Bron1}
reported the direct observation of the stimulated emission of $24.7\ cm^{-1}$
phonons in a three level system of $V^{4+}$ ions in $Al_{2}O_{3}$. Prieur 
{\it et al.} \cite{Prieur1,Prieur2} studied the phonon emission in a two
level system of a glass. These phonons had a wide frequency distribution
around $\omega =0.34\ GHz$. Zavtrak and collaborators \cite
{Zav1,Zav2,Zav3,Zav4,Zav5,Zav6} proposed a device consisting in a dielectric
liquid with small particles or gas bubbles working in such a way similar to
a free-electron laser. That device could produce coherent acoustic phonons
by stimulated emission at a frequency $\omega \approx 2\ KHz$. Stimulated
emission of phonons in an acoustic cavity was obtained experimentally by
Fokker {\it et al.} \cite{Fokker1,Fokker2}. They use the metastable
Zeeman-split doublet in ruby. These phonons have frequencies of about $60\
MHz$.

At present, several kinds of amplified coherent beams of bosons are produced
by stimulated emission. Besides the traditional visible or near infrared
photons (lasers) and microwave photons (masers), other coherent photon beams
are produced. High energy lasing processes like X ray lasers (called also
xasers) \cite{Wang1,Wang2} and gamma ray lasers (GRASERS) \cite{Solem} were
studied.

A very exciting field has been recently opened with the development of atoms
lasers \cite{Ketterle-1,Ketterle-2,Wieman,Yasuda}. Several applications of
these novel devices, like high precision gravimetry, are proposed. Strongly
related, it is the study of vibrational amplification by stimulated emission
of radiation (vaser) \cite{Wallentowitz}.

Kleppner \cite{Kleppner} claims against that complicated jargon of acronyms
and proposes to simplify to atom laser, X ray laser, etc.

In summary, a great effort has been made to produce terahertz phonons using
other mechanisms different of stimulated emission. On the other side,
acoustic phonons have been produced by stimulated emission but at
frequencies lower than $2\ THz$. The DBH resonator proposed here could
produce a continuous beam of terahertz coherent phonons. The ultra short
wavelength of these phonons permits potential applications.

Acoustic imaging is a very well known procedure, mainly in medicine, where
it is exploited the fact that different water contents on several tissues
produce different attenuations for a signal of about $4\ MHz$. Today are
available commercial equipments that can produce three-dimensional images of
very high resolution.

Equipments for acoustic microscopy are less known but they are also
commercially available. The most used technique is Scanning Acoustic
Microscopy (SAM) \cite{Yu} that can attain resolutions higher than $0.4\ \mu
m$, taken with phonons of $2\ GHz$.

Besides the scanning acoustic microscopy, several techniques have been
presented in a recent review \cite{Gilmore}: photoacoustic microscopy (PAM),
scanning electron acoustic microscopy (SEAM) and scanning laser acoustic
microscopy (SLAM). The theory of acoustic imaging was also reviewed,
schematic diagrams of the different microscopes are presented and several
industrial applications are discussed in reference \cite{Gilmore}.

The equipments mentioned above work at up to $10\ GHz$, three orders of
magnitude lower than the frequency of the phonon beam considered in the
present paper. Therefore, they have a limited theoretical resolution given
by its wavelength that is $1000$ times greater than the generated by a DBH
device.

To make the acoustic nanoscopy possible, it is necessary to have a coherent
beam of ultra short wavelength phonons. To obtain images of a small
structure, one would need waves shorter than the size of the details of the
system.

H. Maris \cite{HMaris} describes the process of acoustic nanoscopy by mean
of echoes of ultra-short phonon pulses optically generated. They are
detected by measuring the variation of optical reflectivity due to the
scattered phonon pulses. They work with coherent phonons of about $2\ THz$
like those produced by our device, but they were not produced by a phonon
laser.

Evidently, a laser is not necessary to produce optical images. Nevertheless,
lasers are a very useful tool for imaging and up today they are
indispensable to make holograms. On the other hand, many optical experiments
that can be made with incoherent light are done more easily and more
precisely with the aid of a laser. Therefore many experiments that are made
today with incoherent phonons could be done better with the aid of a phonon
laser.

The phonon laser described in this paper has the wavelength smaller than $25$
{\it \AA }\ (what is about 20 times less than for the device described in 
\cite{Sup}). Therefore, it is possible to use it to see structures of about $%
50\ nm$. A coherent source of X rays of about $20\ nm$ begins to be
available now \cite{Rundquist-1,Rundquist-2}. Even if it will be possible to
make images of structures at the submicron scale by using X rays of a
wavelength of about $25$ {\it \AA }, they would have energies of about $0.5\
KeV$, what could affect the system to be studied.

The device presented here could be used to get a hologram at nano-scale. To
do this, it is possible to build in a {\it GaAs} an acoustic analogous of a
Michelson interferometer. It consists of a thin barrier of {\it AlGaAs}
designed as a semimirror, placed at $45^{o}$ to split the beam and to force
it to pass one half through the part of the sample we are interested in and
the other across a very pure reference {\it GaAs}. After that, we can use
mirrors (made of thick {\it AlGaAs} barriers) to superpose them and record
the interference pattern with a detector. This pattern could be processed in
a computer to get a hologram of the sample showing the three-dimensional
shape of the nanostructure. In a previous paper \cite{jpcm} we described, as
an alternative, the acoustic version of a Sagnac interferometer.

Because of the high sensitivity to all types of lattice defects, the phonon
beam could be used to give information about the inner lattice structure 
\cite{Msall}, \cite{HMaris} and form part of a phonon imaging setup as a
source \cite{Wolfe}.

We will show that the amplitude of the phonon beam is with a great precision
directly proportional to the incoming current. Therefore, the saser
amplitude could be modulated by modulating the amplitude of the current.
This will permit the use of the phonon laser as an information carrier
between circuit components at very short distances. These
``phonoelectronic'' devices could work at smaller distances and lower
energies than optoelectronic devices.

If we achieve to make a ``phonoresist'', i.e., a material with a soft mode
close to the frequency of our phonon laser, the saser could be used for
nanolithography. Due to the short wavelength and low energy of the phonon
beam, ``phonolithography'' would be a better option than the methods
currently in used.

The rest of this paper is organized as follow. The Hamiltonian which
describes our system is presented in Sec. II. The solutions of the
Hamiltonian are obtained is Sec. III. In the Sec. IV, we show the procedure
used to obtain the kinetic equations. We devote Sec. V to present our
results and conclusions.

\section{The Hamiltonian}

To describe our system, we considered a Hamiltonian composed by a single
particle part and one representing the interactions. The single particle
part can be written as

\begin{equation}
{\cal H}_{o} = {\cal H}_{e} + {\cal H}_{1} + {\cal H}_{2} + {\cal H}_{3},
\label{H_o}
\end{equation}

where ${\cal H}_e$ describes the electrons, and ${\cal H}_1$, ${\cal H}_2$, $%
{\cal H}_3$ the $LO$, $\widetilde{LO}$ and $TA$ phonons respectively.

Assuming the grown direction as the $z$ axis, the Hamiltonians ${\cal H}_{1}$%
, ${\cal H}_{2}$, ${\cal H}_{3}$ for the phonons can be written

\begin{equation}
{\cal H}_{1}=\sum_{{\bf q}_{1}}\hbar \omega _{1}b_{1_{{\bf q}_{1}}}^{\dagger
}b_{1_{{\bf q}_{1}}},  \label{H_LO}
\end{equation}

\begin{equation}
{\cal H}_{2}=\sum_{{\bf q}_{2}}\left( \hbar \omega _{2}-i\hbar \kappa
_{2}\right) b_{2_{{\bf q}_{2}}}^{\dagger }b_{2_{{\bf q}_{2}}},  \label{H_LO2}
\end{equation}

\begin{equation}
{\cal H}_{3}=\sum_{{\bf q}_{3}}\left( \hbar \omega _{3}-i\hbar \kappa
_{3}\right) b_{3_{{\bf q}_{3}}}^{\dagger }b_{3_{{\bf q}_{3}}}  \label{H_TA}
\end{equation}

where $\hbar \omega _{1}$, $\hbar \omega _{2}$, $\hbar \omega _{3}$ are the
energies for $LO$, $\widetilde{LO}$ and $TA$ phonons respectively and $b_{i_{%
{\bf q}_{i}}}^{\dagger }$($b_{i_{{\bf q}_{i}}}$) are the creation
(annihilation) operators for phonons with momentum ${\bf q}_{i}$
perpendicular to the $z$ direction. Two imaginary terms $\left( i\hbar
\kappa _{2}\text{, }i\hbar \kappa _{3}\right) $ were introduced to take into
account the decay by anharmonicity of the $\widetilde{LO}$ phonons and the
escape of the $TA$ phonons, respectively.

To describe the electrons, we use a tight binding approach with hoppings $v$
between nearest neighbors. Writing the Hamiltonian in a Wannier basis, in
terms of the electronic creation (annihilation) operators $c_{{\bf r}}^{%
{\cal \dagger }}$($c_{{\bf r}}$) at site ${\bf r}$, we get

\begin{equation}
{\cal H}_{e} =\sum_{{\bf r}}\varepsilon _{{\bf r}}c_{{\bf r}}^{{\cal \dagger 
}}c_{{\bf r}}+v\sum_{<{\bf r},{\bf r}^{\prime }>}\left(c_{{\bf r}}^{{\cal %
\dagger }}c_{{\bf r}^{\prime }}+c_{{\bf r}^{\prime }}^{{\cal \dagger }}c_{%
{\bf r}}\right)  \label{H_e1}
\end{equation}

where ${\bf r}\equiv (l,m,j)$ with $l,$ $m,$ $j=-\infty ,$ $\ldots ,$ $%
\infty $. As the system has not magnetic behavior the subscript $\sigma $
(for spin) is left implicit.

Our system has translational symmetry in the perpendicular directions ($xy$%
). Therefore the Hamiltonian can be uncoupled by expanding the operators $c_{%
{\bf r}}$ in plane waves in these directions

\begin{equation}
c_{{\bf r}}=\sum_{{\bf k}}c_{j{\bf k}}e^{i{\bf k}\cdot {\bf x}_{lm}}.
\label{c}
\end{equation}

We can treat the system as a sum over 1D Hamiltonians for each wave vector $%
{\bf k}$ perpendicular to the current $\left( z\right) $ direction

\begin{equation}
{\cal H}_{e}=\sum_{{\bf k}}\Big\{\sum_{j}\widetilde{\varepsilon }_{j{\bf k}%
}c_{j{\bf k}}^{{\cal \dagger }}c_{j{\bf k}}+v\sum_{<jj^{\prime }>}\left( c_{j%
{\bf k}}^{{\cal \dagger }}c_{j^{\prime }{\bf k}}+c_{j^{\prime }{\bf k}}^{%
{\cal \dagger }}c_{j{\bf k}}\right) \Big\}  \label{H_e2}
\end{equation}

where $\widetilde{\varepsilon }_{j{\bf k}}=\varepsilon _{j{\bf k}}-4v$. The
energies $\widetilde{\varepsilon }_{j{\bf k}}$ are measured from the bottom
of the conduction band of the emitter and $\varepsilon _{j{\bf k}%
}=\varepsilon _{j}+\varepsilon _{{\bf k}}$. The energies $\varepsilon _{j}$
are chosen to describe the energy profile of the DBH. For the sake of
simplicity, we will leave implicit the dependence on ${\bf k}$.

For the $z$ direction, we separate the space in three regions: the
dispersion region and two semi-infinite one-dimensional chains. On the left
we will have planes with energy $\varepsilon _{j}=0$ for $j\leq 0$ and to
the right $\varepsilon _{j}=V$ for $j\geq L+1$, $L$ is the length of the
DBH. The corresponding eigenstates of these two regions are planes waves. If
we disconnect the DBH from the left and right chains we get for the
dispersion region the profile of an infinite (even not rectangular) well as
it is shown in figure \ref{well-Fig2}.

For the dispersion region, we diagonalize a three-diagonal matrix of order $%
L $ corresponding to the profile showed above getting the eigenvalues $%
\varepsilon _{m}$ and the eigenvectors $|m\rangle $ of equation

\begin{equation}
{\cal H}_{ez}^{\prime }|m\rangle =\varepsilon _{m}|m\rangle .  \label{Hez}
\end{equation}

Here, ${\cal H}_{ez}^{\prime }$ is the part of ${\cal H}_{ez}$ that goes
from the beginning of the left barrier to the end of the right one. Written
in the basis of planes this is

\begin{equation}
{\cal H}_{ez}^{\prime }=\left( 
\begin{array}{cccccc}
\widetilde{\varepsilon }_{1} & v & 0 &  &  &  \\ 
v & \widetilde{\varepsilon }_{2} & v &  &  &  \\ 
0 & v & \widetilde{\varepsilon }_{3} & \ddots &  &  \\ 
&  & \ddots & \ddots &  &  \\ 
&  &  &  &  & v \\ 
&  &  &  & v & \widetilde{\varepsilon }_{L}
\end{array}
\right) ,  \label{Matrix}
\end{equation}

where $\widetilde{\varepsilon }_{j}=\varepsilon _{j}-2v$.

After the diagonalization we get $L$ discrete levels. We label these levels
through the index $m$ as $m=0$, $1^{\prime }$, $2^{\prime }\ldots $. For the
right chain we rename the planes, from $L+1$, $L+2$, $\ldots $ to $1$, $2$, $%
\ldots $; and for the left one, from $0$, $-1$, $-2$,$\ldots $, $-\infty $
to $\overline{1}$, $\overline{2}$, $\ldots $, $\overline{\infty }$.
Therefore we get a new picture where we have two semi-infinite chains, the
first one corresponding to the planes $j=\overline{\infty }$, $\ldots $, $%
\overline{1}$ and the other one corresponding to $j=1$, $\ldots $, $\infty $%
. Among the $L$ levels obtained from the diagonalization of the matrix (\ref
{Matrix}), only the two states with lowest energies will participate
significantly in the electronic transport. The others are far over the Fermi
level and their effects will be considered later.

To connect the DBH with the left and right chains we compute the matrix
elements

\begin{equation}
v_{jm}=\langle j|{\cal H}_{ez}|m\rangle .  \label{vjm}
\end{equation}

The electronic Hamiltonian is

\begin{eqnarray}
{\cal H}_{e} &=&\sum_{j}\varepsilon _{j}c_{j}^{{\cal \dagger }%
}c_{j}+\sum_{m}\varepsilon _{m}c_{m}^{{\cal \dagger }}c_{m}+\sum_{j\neq 
\overline{1},m}v\left( c_{j}^{{\cal \dagger }}c_{j+1}+c_{j+1}^{{\cal \dagger 
}}c_{j}\right) +  \nonumber \\
&&\sum_{m}\Big\{v_{\overline{1}m}\left( c_{\overline{1}}^{{\cal \dagger }%
}c_{m}+c_{m}^{{\cal \dagger }}c_{\overline{1}}\right) +{}v_{m1}\left( c_{m}^{%
{\cal \dagger }}c_{1}+c_{1}^{{\cal \dagger }}c_{m}\right) \Big\}.
\label{H_e}
\end{eqnarray}

A diagram representing the electronic Hamiltonian (\ref{H_e}) is shown in
figure \ref{diagram_He}.

As it is well known \cite{Haken1}, the dominant dispersion process in polar
semiconductors is due to the coupling between electrons and $LO$ phonons.

The interaction Hamiltonian ${\cal H}_{int}$ has three terms

\begin{equation}
{\cal H}_{int}={\cal H}_{e-e}+{\cal H}_{e-ph}+{\cal H}_{ph-ph}  \label{H_int}
\end{equation}

where ${\cal H}_{e-e}$, ${\cal H}_{e-ph}$ and ${\cal H}_{ph-ph}$ describe
the electron-electron, the electron-phonon, and the phonon-phonon
interactions respectively.

${\cal H}_{e-e}$ is taken in the Hartree approximation. That means that the
energies $\varepsilon _{m}$ of the levels in the well depend on the
accumulated charge.

The electron-phonon interaction Hamiltonian has the form,

\begin{equation}
{\cal H}_{e-ph}=\sum_{{\bf k},{\bf q}_{1}}g_{_{{\bf kq}_{1}}}(c_{0_{{\bf k-q}%
_{1}}}^{{\cal \dagger }}c_{1^{\prime }{\bf k}}b_{1_{{\bf q}_{1}}}^{{\cal %
\dagger }}+c_{1^{\prime }{\bf k}}^{{\cal \dagger }}c_{0_{{\bf k-q}%
_{1}}}b_{1_{{\bf q}_{1}}}).  \label{H_e-ph1}
\end{equation}

In order to recover the translational symmetry in the $xy$ plane we take the
average on ${\bf q}_{1}$ \cite{jpcm}

\begin{equation}
{\cal H}_{e-ph}=\sum_{{\bf k}}g_{_{{\bf k}}}\left( c_{0_{{\bf k}}}^{{\cal %
\dagger }}c_{1^{\prime }{\bf k}}b_{1_{{\bf k}}}^{{\cal \dagger }%
}+c_{1^{\prime }{\bf k}}^{{\cal \dagger }}c_{0_{{\bf k}}}b_{1_{{\bf k}%
}}\right)  \label{H_e-ph}
\end{equation}

where $g_{{\bf k}}$ measure the effective strength of the interaction and
was calculated as in \cite{GWeber}.

The phonon-phonon interaction is given by \cite{Vall}

\begin{equation}
{\cal H}_{ph-ph}=\sum_{{\bf q}_{1}}\gamma _{{\bf q}_{1}}(b_{1_{{\bf q}%
_{1}}}b_{2_{{\bf q}_{1}}}^{{\cal \dagger }}b_{3}^{{\cal \dagger }}+b_{1_{%
{\bf q}_{1}}}^{{\cal \dagger }}b_{2_{{\bf q}_{1}}}b_{3}).  \label{H_ph-ph}
\end{equation}

This Hamiltonian describes the decay $LO\rightarrow \widetilde{LO}+TA$ and
its inverse process (recombination).

As in the {\it GaAs} the $TA$ phonons are emitted fundamentally in the [$111$%
] direction \cite{Vall}, we considered here a device that was grown in this
direction. The inside barrier walls act as mirrors of a Fabry-Perot
interferometer selecting discrete cavity modes. This mirrors also serve for
a selection of the saser modes: those running in the grown direction are
reflected several times between the mirrors and stay longer in the well,
while all others are lost.

With the increasing of the applied bias $V$, from a certain value, called
saser threshold, the $TA$ phonons associated with the cavity resonant mode,
i.e., $q_{3}=0$, stimulate strongly the $LO$ phonon decay and their number
grow sharply. Once this mode is sufficiently populated, the probability of
stimulated emission for that mode exceeds the sum of the emission
probabilities for all other modes. Thus, the $TA$ phonons emitted in this
mode will slave the other ones, giving rise to a coherent acoustic phonon
beam. This process is similar to what happens in a laser. Due to the
momentum conservation ${\bf q}_{1}={\bf q}_{2}+{\bf q}_{3}$ and because of $%
{\bf q}_{3}=0$ \cite{ssc1} we get ${\bf q}_{2}={\bf q}_{1}$.

The parameter $\gamma _{{\bf q}_{1}}$ appearing in (\ref{H_ph-ph})
represents the intensity of the interaction and was estimated from \cite
{Vall}, \cite{Klemens} and \cite{Ziman}.

Finally, the total Hamiltonian is the sum of both terms: the single particle
and the interaction parts

\begin{equation}
{\cal H}=\sum_{{\bf k}}{\cal H}_{o}\left( {\bf k}\right) +{\cal H}%
_{int}\left( {\bf k}\right)  \label{H}
\end{equation}

where we had made the approximation that the momentum ${\bf q}_{1}$
(appearing in expressions (\ref{H_LO}), (\ref{H_LO2}) and (\ref{H_ph-ph}))
varies in the same range as ${\bf k}$. We can see at expression (\ref{H})
that the Hamiltonian became, in this approximation, fully uncoupled for each 
${\bf k}$. This simplifies strongly the obtention of its solutions.

\section{The resolution of the Hamiltonian}

In order to solve de Hamiltonian described in the former section we
introduce the operators

\begin{equation}
{\cal O}_{j{\bf k}}^{n_{1},n_{2},n_{3}}=c_{j{\bf k}}\frac{b_{1{\bf k}%
}^{n_{1}}b_{2{\bf k}}^{n_{2}}b_{3}^{n_{3}}}{\sqrt{n_{1}!}\sqrt{n_{2}!}\sqrt{%
n_{3}!}}.  \label{O}
\end{equation}

The operators ${\cal O}_{j{\bf k}}^{{n_{1},n_{2},n_{3}}^{\dagger }}$ creates
the eigenstates of ${\cal H}_{o}$ (without the hopping terms)

\begin{equation}
|j{\bf k}n_{1}n_{2}n_{3}\rangle \equiv {\cal O}_{j{\bf k}}^{{%
n_{1},n_{2},n_{3}}^{\dagger }}|0\rangle  \label{Coherent}
\end{equation}

formed by an electron at plane $j$ and $n_{1}$ $LO$ phonons, $n_{2}$ $%
\widetilde{LO}$ phonons with parallel momentum ${\bf k}$; and $n_{3}$ $TA$
phonons.

As it was done before, we will leave implicit the dependence on ${\bf k}$.

Using the operators (\ref{O}) and the Hamiltonian (\ref{H}) we can calculate
their equations of motion. In order to simplify the notation we call ${\bf n}
$ for the set of $\left\{ n_{1},n_{2},n_{3}\right\} $ phonons

\begin{equation}
i\hbar \frac{d{\cal O}_{j}^{{\bf n}}}{dt}=\left[ {\cal O}_{j}^{{\bf n}},%
{\cal H}\right] .  \label{O_mov}
\end{equation}

By expanding the eigenstates $|\Phi \rangle $ of ${\cal H}$ in the
eigenstates of ${\cal H}_o$

\begin{equation}
|\Phi \rangle =\sum_{j,{\bf n}}p_j^{{\bf n}}|j{\bf n\rangle },  \label{Fi}
\end{equation}

and due to the orthogonality of $|j{\bf n\rangle ,}$ the amplitudes can be
calculated as

\begin{equation}
p_{j}^{{\bf n}}=\langle jn|\Phi \rangle {\bf =\langle }0{\bf |}{\cal O}_{j}^{%
{\bf n}}|\Phi \rangle {\bf .}
\end{equation}

Making use of the equations turning out from (\ref{O_mov}) and the previous
definition for the amplitudes, we get for $j\neq \overline{1},1,m\quad
\left( m=0,1^{\prime },2^{\prime },3^{\prime }\ldots L\right) $

\begin{equation}
i\hbar {{\frac{dp_{j}^{{\bf n}}}{dt}}}=\varepsilon _{j}^{{\bf n}}p_{j}^{{\bf %
n}}+\Upsilon _{j}+vp_{j+1}^{{\bf n}}+vp_{j-1}^{{\bf n}}  \label{dPj}
\end{equation}

and for $j=\overline{1},1,m$

\begin{mathletters}
\label{dP}
\begin{eqnarray}
i\hbar {{\frac{dp_{\overline{1}}^{{\bf n}}}{dt}}} &=&\varepsilon _{\overline{%
1}}^{{\bf n}}p_{\overline{1}}^{{\bf n}}+\Upsilon _{\overline{1}}+v_{%
\overline{1}1^{\prime }}p_{1^{\prime }}^{{\bf n}}+v_{\overline{1}0}p_{0}^{%
{\bf n}}+vp_{\overline{2}}^{{\bf n}}+\sum_{m\geq 2^{\prime }}v_{\overline{1}%
m}p_{m}^{{\bf n}},  \label{dPm1} \\
i\hbar {{\frac{dp_{0}^{{\bf n}}}{dt}}} &=&\varepsilon _{0}^{{\bf n}}p_{0}^{%
{\bf n}}+g\sqrt{n_{1}}p_{1^{\prime }}^{{\bf n^{-}}}+\Upsilon _{0}+v_{%
\overline{1}0}p_{\overline{1}}^{{\bf n}}+v_{01}p_{1}^{{\bf n}},  \label{dP0}
\\
i\hbar {{\frac{dp_{1^{\prime }}^{{\bf n}}}{dt}}} &=&\varepsilon _{1^{\prime
}}^{{\bf n}}p_{1^{\prime }}^{{\bf n}}+g\sqrt{n_{1}+1}p_{0}^{{\bf n^{+}}%
}+\Upsilon _{1^{\prime }}+v_{\overline{1}1^{\prime }}p_{\overline{1}}^{{\bf n%
}}+v_{1^{\prime }1}p_{1}^{{\bf n}},  \label{dPm0} \\
i\hbar {{\frac{dp_{1}^{{\bf n}}}{dt}}} &=&\varepsilon _{1}^{{\bf n}}p_{1}^{%
{\bf n}}+\Upsilon _{1}+v_{1^{\prime }1}p_{1^{\prime }}^{{\bf n}%
}+v_{01}p_{0}^{{\bf n}}+vp_{2}^{{\bf n}}+\sum_{m\geq 2^{\prime
}}v_{1m}p_{m}^{{\bf n}},  \label{dP1} \\
i\hbar {{\frac{dp_{m}^{{\bf n}}}{dt}}} &=&\varepsilon _{m}^{{\bf n}}p_{m}^{%
{\bf n}}+v_{1m}p_{1}^{{\bf n}}+v_{\overline{1}m}p_{\overline{1}}^{{\bf n}%
}+\Upsilon _{m},\quad m\geq 2^{\prime }  \label{dPm}
\end{eqnarray}

where

\end{mathletters}
\begin{eqnarray*}
\varepsilon _{j}^{{\bf n}} &=&\varepsilon _{j}+n_{1}\hbar \omega
_{1}+n_{2}\left( \hbar \omega _{2}-i\hbar \kappa _{2}\right) + n_{3}\left(
\hbar \omega _{3}-i\hbar \kappa _{3}\right) , \\
{\bf n}^{-} &\equiv &\left\{ n_{1}-1,n_{2},n_{3}\right\} , \\
{\bf n}^{+} &\equiv &\left\{ n_{1}+1,n_{2},n_{3}\right\} , \\
\Upsilon _{j} &=&\gamma \left\{ \sqrt{N^{-}}p_{j}^{{\bf n^{\prime }}}+\sqrt{%
N^{+}}p_{j}^{{\bf n^{\prime \prime }}}\right\} , \\
{\bf n}^{\prime } &\equiv &\left\{ n_{1}-1,n_{2}+1,n_{3}+1\right\} , \\
{\bf n}^{\prime \prime } &\equiv &\left\{ n_{1}+1,n_{2}-1,n_{3}-1\right\} ,
\\
N^{-} &=&n_{1}\left( n_{2}+1\right) \left( n_{3}+1\right) , \\
N^{+} &=&\left( n_{1}+1\right) n_{2}n_{3}.
\end{eqnarray*}

The amplitudes $p_{j}^{{\bf n}}$ are time dependent. We transform them

\begin{equation}
p_{j}^{{\bf n}}\left( t\right) ={{\frac{1}{\sqrt{2\pi }}}}\int_{-\infty
}^{\infty }d\omega a_{j}^{{\bf n}}\left( \omega \right) e^{-i\omega t}.
\label{Fourier1}
\end{equation}

By doing that, we get the following set of equations

\begin{mathletters}
\label{1Amplitudes}
\begin{eqnarray}
&&\hbar \omega a_{j}^{{\bf n}}=\varepsilon _{j}^{{\bf n}}a_{j}^{{\bf n}%
}+\Gamma _{j}+va_{j+1}^{{\bf n}}+va_{j-1}^{{\bf n}},\quad \text{for }j\neq 
\overline{1},1,m \\
&&\hbar \omega a_{\overline{1}}^{{\bf n}}=\varepsilon _{\overline{1}}^{{\bf n%
}}a_{\overline{1}}^{{\bf n}}+\Gamma _{\overline{1}}+va_{\overline{2}}^{{\bf n%
}}+v_{\overline{1}1^{\prime }}a_{1^{\prime }}^{{\bf n}}+v_{\overline{1}%
0}a_{0}^{{\bf n}}+\sum_{m\geq 2^{\prime }}v_{\overline{1}m}a_{m}^{{\bf n}},
\label{1Am1} \\
&&\hbar \omega a_{0}^{{\bf n}}=\varepsilon _{0}^{{\bf n}}a_{0}^{{\bf n}}+g%
\sqrt{n_{1}}a_{1^{\prime }}^{{\bf n}^{-}}+\Gamma _{0}+v_{\overline{1}0}a_{%
\overline{1}}^{{\bf n}}+v_{01}a_{1}^{{\bf n}},  \label{1A0} \\
&&\hbar \omega a_{1^{\prime }}^{{\bf n}}=\varepsilon _{1^{\prime }}^{{\bf n}%
}a_{1^{\prime }}^{{\bf n}}+g\sqrt{n_{1}+1}a_{0}^{{\bf n^{+}}}+\Gamma
_{1^{\prime }}+v_{\overline{1}1^{\prime }}a_{\overline{1}}^{{\bf n}%
}+v_{1^{\prime }1}a_{1}^{{\bf n}},  \label{1Am0} \\
&&\hbar \omega a_{1}^{{\bf n}}=\varepsilon _{1}^{{\bf n}}a_{1}^{{\bf n}%
}+\Gamma _{1}+va_{2}^{{\bf n}}+v_{1^{\prime }1}a_{1^{\prime }}^{{\bf n}%
}+v_{01}a_{0}^{{\bf n}}+\sum_{m\geq 2^{\prime }}v_{1m}a_{m}^{{\bf n}},
\label{1A1} \\
&&\hbar \omega a_{m}^{{\bf n}}=\varepsilon _{m}^{{\bf n}}a_{m}^{{\bf n}%
}+\Gamma _{m}+v_{\overline{1}m}a_{\overline{1}}^{{\bf n}}+v_{1m}a_{1}^{{\bf n%
}},\quad m\geq 2^{\prime }.  \label{1Am}
\end{eqnarray}

Here $\Gamma _{j}$ is the Fourier transform of $\Upsilon _{j}$.

In order to obtain the solutions of (\ref{1Amplitudes}), we considered only
three channels through which the electrons can tunnel. Such channels are the
only that are between the bottom of the conduction band ($\varepsilon =0$)
and the Fermi energy $\varepsilon _{F}^{L}$ at the emitter, near the
resonant condition. Therefore, they are the only ones that contribute
significantly to the current. These channels are: one electron through the
excited level and no phonons (channel $0$, i.e., ${\bf n}=\{0,0,0\}$); one
electron through the ground level and one $LO$ phonon (channel $100$) and
one electron through the ground level and a pair $\widetilde{LO}-TA$ phonons
respectively (channel $011$).

To solve the system above, we drop the imaginary part of $\varepsilon _{j}^{%
{\bf n}}$. This cope with the problems caused by the non-Hermitian
Hamiltonian.

We have to note that the amplitudes $a_{j}^{0}$ (for $j\geq 1$ or $j\leq 
\overline{1}$) remain uncoupled (this tunneling channel does not mix with
any other). On the other hand, the other two channels are coupled due to the
ph-ph interaction. To uncouple them we propose the following transformation
for the amplitudes

\end{mathletters}
\begin{eqnarray}
a_{j}^{100} &=&{{\frac{1}{\sqrt{2}}}}\left( a_{j}^{+}+a_{j}^{-}\right) ,
\label{Aj100} \\
a_{j}^{011} &=&{{\frac{1}{\sqrt{2}}}}\left( a_{j}^{+}-a_{j}^{-}\right) .
\label{Aj011}
\end{eqnarray}

Making the above transformations for the amplitudes $a_{j}^{100}$ and $%
a_{j}^{011}$ we get two new uncoupled channels (called channels $+$ and $-$).

Considering only the channels $0$ and $\pm $, the system (\ref{1Amplitudes})
can be rewritten as

\begin{mathletters}
\label{AllEq}
\begin{eqnarray}
&&va_{j-1}^{0}+\left( \varepsilon _{j}^{0}-\hbar \omega \right)
a_{j}^{0}+va_{j+1}^{0}=0  \label{Eq1} \\
&&va_{j-1}^{\pm }+\left( \varepsilon _{j}^{100}\pm \gamma -\hbar \omega
\right) a_{j}^{\pm }+va_{j+1}^{\pm }=0\quad \text{for }j\neq \overline{1}%
,m,1;  \label{Eq2} \\
&&va_{\overline{2}}^{0}+\left( \varepsilon _{\overline{1}}^{0}-\hbar \omega
\right) a_{\overline{1}}^{0}+v_{\overline{1}1^{\prime }}a_{1^{\prime
}}^{0}+\sum_{m\neq 1^{\prime }}v_{\overline{1}m}a_{m}^{0}=0  \label{Eq3} \\
&&va_{\overline{2}}^{\pm }+\left( \varepsilon _{\overline{1}}^{100}\pm
\gamma -\hbar \omega \right) a_{\overline{1}}^{\pm }+v_{\overline{1}%
0}a_{0}^{\pm }+\sum_{m\neq 0}v_{\overline{1}m}a_{m}^{\pm }=0\qquad 
\label{Eq4} \\
&&v_{1^{\prime }1}a_{1^{\prime }}^{0}+\sum_{m\neq 1^{\prime
}}v_{m1}a_{m}^{0}+\left( \varepsilon _{1}^{0}-\hbar \omega \right)
a_{1}^{0}+va_{2}^{0}=0\qquad   \label{Eq5} \\
&&v_{01}a_{0}^{\pm }+\sum_{m\neq 1^{\prime }}v_{m1}a_{m}^{\pm }+\left(
\varepsilon _{1}^{100}\pm \gamma -\hbar \omega \right) a_{1}^{\pm
}+va_{2}^{\pm }=0  \label{Eq6} \\
&&v_{\overline{1}m}a_{\overline{1}}^{\pm }+\left( \varepsilon _{m}^{100}\pm
\gamma -\hbar \omega \right) a_{m}^{\pm }+v_{m1}a_{1}^{\pm }=0\quad \text{%
for }m\neq 0;  \label{Eq7} \\
&&v_{\overline{1}m}a_{\overline{1}}^{0}+\left( \varepsilon _{m}^{0}-\hbar
\omega \right) a_{m}^{0}+v_{m1}a_{1}^{0}=0\quad \text{for }m\neq 1^{\prime };
\label{Eq8} \\
&&v_{\overline{1}0}a_{\overline{1}}^{\pm }+\left( \varepsilon _{0}^{100}\pm
\gamma -\hbar \omega \right) a_{0}^{\pm }+{{\frac{g}{\sqrt{2}}}}a_{1^{\prime
}}^{0}+v_{01}a_{1}^{\pm }=0  \label{Eq9} \\
&&v_{\overline{1}1^{\prime }}a_{\overline{1}}^{0}+\left( \varepsilon
_{1^{\prime }}^{0}-\hbar \omega \right) a_{1^{\prime }}^{0}+{{\frac{g}{\sqrt{%
2}}}}a_{0}^{+}+{{\frac{g}{\sqrt{2}}}}a_{0}^{-}+v_{1^{\prime }1}a_{1}^{0}=0
\label{Eq10}
\end{eqnarray}

The amplitudes $a_{m}^{\pm }$ and $a_{m}^{0}$ appearing in equations (\ref
{AllEq}) take into account the electronic energy levels different from the
fundamental and excited one. To eliminate these amplitudes (the fundamental
and excited levels are the only ones that participate directly in the $LO$
phonon generation) we use the equations (\ref{Eq3})-(\ref{Eq8}). The
resultant system is

\end{mathletters}
\begin{mathletters}
\label{Amplitudes}
\begin{eqnarray}
&&va_{j-1}^{0}+\left( \varepsilon _{j}^{0}-\hbar \omega \right)
a_{j}^{0}+va_{j+1}^{0}=0  \label{Aj_0} \\
&&va_{j-1}^{\pm }+\left( \varepsilon _{j}^{100}\pm \gamma -\hbar \omega
\right) a_{j}^{\pm }+va_{j+1}^{\pm }=0\quad \text{for\quad }j\neq \overline{1%
},m,1;  \label{Aj_+-} \\
&&va_{\overline{2}}^{0}+\left( \widetilde{\varepsilon }_{\overline{1}%
}^{0}-\hbar \omega \right) a_{\overline{1}}^{0}+v_{\overline{1}1^{\prime
}}a_{1^{\prime }}^{0}+\widetilde{v}_{\overline{1}1}^{0}a_{1}^{0}=0
\label{Am1_0} \\
&&va_{\overline{2}}^{\pm }+\left( \widetilde{\varepsilon }_{\overline{1}%
}^{\pm }\pm \gamma -\hbar \omega \right) a_{\overline{1}}^{\pm }+v_{%
\overline{1}0}a_{0}^{\pm }+\widetilde{v}_{\overline{1}1}^{\pm }a_{1}^{\pm }=0
\label{Am1_+-} \\
&&\widetilde{v}_{\overline{1}1}^{0}a_{\overline{1}}^{0}+v_{1^{\prime
}1}a_{1^{\prime }}^{0}+\left( \widetilde{\varepsilon }_{1}^{0}-\hbar \omega
\right) a_{1}^{0}+va_{2}^{0}=0  \label{A1_0} \\
&&\widetilde{v}_{\overline{1}1}^{\pm }a_{\overline{1}}^{\pm
}+v_{01}a_{0}^{\pm }+\left( \widetilde{\varepsilon }_{1}^{\pm }\pm \gamma
-\hbar \omega \right) a_{1}^{\pm }+va_{2}^{\pm }=0  \label{A1_+-} \\
&&v_{\overline{1}1^{\prime }}a_{\overline{1}}^{0}+\left( \varepsilon
_{1^{\prime }}^{0}-\hbar \omega \right) a_{1^{\prime }}^{0}+{{\frac{g}{\sqrt{%
2}}}}a_{0}^{+}+{{\frac{g}{\sqrt{2}}}}a_{0}^{-}+v_{1^{\prime }1}a_{1}^{0}=0
\label{A_e} \\
&&v_{\overline{1}0}a_{\overline{1}}^{\pm }+\left( \varepsilon _{0}^{100}\pm
\gamma -\hbar \omega \right) a_{0}^{\pm }+{{\frac{g}{\sqrt{2}}}}a_{1^{\prime
}}^{0}+v_{01}a_{1}^{\pm }=0  \label{A_f}
\end{eqnarray}

where

\end{mathletters}
\begin{equation}
\widetilde{v}_{\overline{1}1}^{0}=\sum_{m\neq 1^{\prime }}{{\frac{v_{%
\overline{1}m}v_{m1}}{\left( \hbar \omega -\varepsilon _{m}^{0}\right) }}}
\label{hopp1}
\end{equation}
and

\begin{equation}
\widetilde{v}_{\overline{1}1}^{\pm }=\sum_{m\neq 0}{{\frac{v_{\overline{1}%
m}v_{m1}}{\left( \hbar \omega -\varepsilon _{m}^{100}\mp \gamma \right) }}}
\label{hopp2}
\end{equation}

are renormalized hoppings joining sites $j=\overline{1},1$, and

\begin{mathletters}
\label{rEnergies}
\begin{eqnarray}
\widetilde{\varepsilon }_{\overline{1}}^{0} &=&\varepsilon _{\overline{1}%
}^{0}+\Sigma _{\overline{1}}^{0}  \label{e1} \\
\widetilde{\varepsilon }_{\overline{1}}^{\pm } &=&\varepsilon _{\overline{1}%
}^{100}+\Sigma _{\overline{1}}^{\pm }  \label{e2} \\
\widetilde{\varepsilon }_{1}^{0} &=&\varepsilon _{1}^{0}+{\Sigma }_{1}^{0}
\label{e3} \\
\widetilde{\varepsilon }_{1}^{\pm } &=&\varepsilon _{1}^{100}+\Sigma
_{1}^{\pm }  \label{e4}
\end{eqnarray}

are renormalized energies at the sites $j=\overline{1},1$ for each channel,
where

\end{mathletters}
\begin{mathletters}
\label{rSigmas}
\begin{eqnarray}
\Sigma _{\overline{1}}^{0} &=&\sum_{m\neq 1^{\prime }}{{\frac{v_{\overline{1}%
m}^{2}}{\left( \hbar \omega -\varepsilon _{m}^{0}\right) }}} \\
\Sigma _{\overline{1}}^{\pm } &=&\sum_{m\neq 0}{{\frac{v_{\overline{1}m}^{2}%
}{\left( \hbar \omega -\varepsilon _{m}^{100}\mp \gamma \right) }}} \\
{\Sigma }_{1}^{0} &=&\sum_{m\neq 1^{\prime }}{{\frac{v_{m1}^{2}}{\left(
\hbar \omega -\varepsilon _{m}^{0}\right) }}} \\
\Sigma _{1}^{\pm } &=&\sum_{m\neq 0}{{\frac{v_{m1}^{2}}{\left( \hbar \omega
-\varepsilon _{m}^{100}\mp \gamma \right) }.}}
\end{eqnarray}

A diagram representing equations (\ref{Amplitudes}) is shown in figure \ref
{Diag1-Fig3}.

To solve the system (\ref{Amplitudes}) we proceed as follows.

For the channel $0$ and $\pm $, in the region $j\geq 1$ the solutions of (%
\ref{Aj_0}) and (\ref{Aj_+-}) are respectively

\end{mathletters}
\begin{eqnarray}
a_{j}^{0} &=&A_{T}^{0}e^{ik_{0}^{^{\prime }}z_{j}}  \label{At000} \\
a_{j}^{\pm } &=&A_{T}^{{\bf \pm }}e^{ik^{^{\pm }}z_{j}}  \label{At_+-}
\end{eqnarray}

where $z_{j}$ is the coordinate of the plane $j$ ($z_{j}=ja$; $a$ is the
lattice parameter), $A_{T}^{0}$, $A_{T}^{{\bf \pm }}$ are the amplitudes of
the transmitted waves, and $k_{0}^{^{\prime }}$ and $k^{^{\pm }}$ are the $z$
component of the waves vectors that fulfills the dispersion relation

\begin{eqnarray}
\hbar \omega &=&\varepsilon _{1}^{0}+2v\cos \left( k_{0}^{^{\prime
}}a\right) ,  \label{Disp_k000} \\
\hbar \omega &=&\varepsilon _{1}^{100}+2v\cos \left( k^{\pm }a\right) \pm
\gamma  \label{Disp_k+}
\end{eqnarray}

We did not consider the terms $e^{-ik_{{\bf n}}^{^{\prime }}z_{j}}$ because
we assume that there are no electrons coming in from the right of the DBH.

In the region $~j\leq -1$ we have two kinds of solutions. The first one is
obtained for the excited level that is above the bottom of the conduction
band, in this case the solution is

\begin{equation}
a_{j}^{0}=A_{R}e^{-ik_{0}z_{j}}+A_{I}e^{ik_{0}z_{j}}  \label{Ar}
\end{equation}

where $A_{R}$ and $A_{I}$ are the amplitudes of the reflected and incident
waves and $k_{0}$ satisfies

\begin{equation}
\hbar \omega =\varepsilon _{\overline{1}}^{0}+2v\cos \left( k_{0}a\right) .
\label{Disp_k}
\end{equation}

The second kind of solution is obtained for the ground state that is below
the bottom of the conduction band then, in this case, the solutions are
evanescent modes with amplitudes $A_{E}^{+}$ and $A_{E}^{-}$

\begin{equation}
a_{j}^{\pm }=A_{E}^{{\bf \pm }}e^{\kappa ^{^{\pm }}z_{j}}  \label{Ae_+-}
\end{equation}

where the $\kappa ^{\pm }$ are positive and fulfill the relations

\begin{equation}
\hbar \omega =\varepsilon _{\overline{1}}^{100}+2v\cosh \left( \kappa ^{\pm
}a\right) \pm \gamma .  \label{Disp_kappa}
\end{equation}

After replacing (\ref{At000}), (\ref{At_+-}), (\ref{Ar}), and (\ref{Ae_+-})
in (\ref{Amplitudes}) we get a system of nine equations with the following
nine unknown quantities

\[
a_{1^{\prime }}^{0},\ a_{0}^{+},\ a_{0}^{-},\ A_{R},\ A_{T}^{0},\
A_{E}^{-},\ A_{E}^{+},\ A_{T}^{-}\ \text{and }A_{T}^{+}. 
\]

The solution of this system permit to calculate the electronic current
taking into account the e-ph and ph-ph interactions and to obtain a set of
five kinetic equations as it is shown in the next section.

\section{The kinetic equations}

To get the kinetic equations that describe the population of electrons and
phonons we need to obtain the equations of motion for the electron number
operators $c_{j}^{{\cal \dagger }}c_{j}$ and the phonon number operators $%
b_{i}^{{\cal \dagger }}b_{i}$

\begin{equation}
i\hbar \frac{d}{dt}\left( c_{j}^{{\cal \dagger }}c_{j}\right) =\left[ c_{j}^{%
{\cal \dagger }}c_{j},{\cal H}\right] ,\ i\hbar \frac{d}{dt}\left( b_{i}^{%
{\cal \dagger }}b_{i}\right) =\left[ b_{i}^{{\cal \dagger }}b_{i},{\cal H}%
\right] .  \label{Eq_mov}
\end{equation}

We define the average populations for electrons in the excited and ground
states as 
\[
\overline{n_{j}}=\sum_{occ}\langle \Phi |c_{j}^{{\cal \dagger }}c_{j}|\Phi
\rangle \text{ for }j=0,\ 1^{\prime }. 
\]

From the definition (\ref{Fi}) of $|\Phi \rangle $, the electrons average
populations can be calculated as

\begin{equation}
\overline{n_{j}}=\sum_{{\bf n},occ}\left| p_{j}^{{\bf n}}\right| ^{2}
\label{pop1}
\end{equation}

For the $LO$, $\widetilde{LO}$, and $TA$ phonons, the average populations
are 
\begin{equation}
\overline{n_{i}}=\sum_{occ}\langle \Phi |b_{i}^{{\cal \dagger }}b_{i}|\Phi
\rangle =\sum_{{\bf n},occ,j}\left| p_{j}^{{\bf n}}\right| ^{2}n_{i}\text{%
\quad for }i=1,2,3  \label{pop2}
\end{equation}

Using the equations (\ref{Eq_mov}) we get the set of quantum kinetic
equations

\begin{mathletters}
\label{kin_eq}
\begin{eqnarray}
\frac{d\overline{n_{1^{\prime }}}}{dt} &=&G_{1^{\prime }}^{I}-G_{1^{\prime
}}^{O}-G_{1^{\prime }}^{E}  \label{Kin_eq_1} \\
\frac{d\overline{n_{0}}}{dt} &=&G_{0}^{I}-G_{0}^{O}+G_{1^{\prime }}^{E}
\label{Kin_eq_2} \\
\frac{d\overline{n_{1}}}{dt} &=&G_{1^{\prime }}^{E}-G_{LO}^{E}
\label{Kin_eq_3} \\
\frac{d\overline{n_{2}}}{dt} &=&G_{LO}^{E}-D_{\widetilde{LO}}
\label{Kin_eq_4} \\
\frac{d\overline{n_{3}}}{dt} &=&G_{LO}^{E}-E_{TA},  \label{Kin_eq_5}
\end{eqnarray}

where

\end{mathletters}
\begin{equation}
G_{1^{\prime }}^{I}=\frac{2}{\hbar }\sum_{{\bf n},occ}v_{\overline{1}%
1^{\prime }}%
\mathop{\rm Im}%
\left\{ p_{1^{\prime }}^{{\bf n^{*}}}p_{\overline{1}}^{{\bf n}}\right\} ,
\label{iG0`}
\end{equation}

is the electronic input rate from the emitter to the excited level,

\begin{equation}
G_{1^{\prime }}^{O}=\frac{2}{\hbar }\sum_{{\bf n},occ}v_{1^{\prime }1}%
\mathop{\rm Im}%
\left\{ p_{1}^{{\bf n^{*}}}p_{1^{\prime }}^{{\bf n}}\right\} ,  \label{oG0`}
\end{equation}

is the output rate from there to the collector.

\begin{equation}
G_{1^{\prime }}^{E}=\frac{2}{\hbar }\sum_{{\bf n},occ}g%
\mathop{\rm Im}%
\left\{ p_{0}^{{\bf n}^{+*}}p_{1^{\prime }}^{{\bf n}}\right\} \sqrt{n_{1}+1},
\label{eG0`}
\end{equation}

represents the net balance between emission and absorption of $LO$ phonons
via electron transitions,

\begin{equation}
G_{0}^{I}=\frac{2}{\hbar }\sum_{{\bf n},occ}v_{\overline{1}0}%
\mathop{\rm Im}%
\left\{ p_{0}^{{\bf n^{*}}}p_{\overline{1}}^{{\bf n}}\right\} ,  \label{iG0}
\end{equation}

is the input rate from the emitter directly to the ground state and

\begin{equation}
G_{0}^{O}=\frac{2}{\hbar }\sum_{{\bf n},occ}v_{01}%
\mathop{\rm Im}%
\left\{ p_{1}^{{\bf n^{*}}}p_{0}^{{\bf n}}\right\} .  \label{oG0}
\end{equation}

stands for the output rate from the ground state to the collector.

The balance between the decay of one phonon $LO$ by emission of a pair of $%
\widetilde{LO}$ and $TA$ phonons, and its inverse process (i.e.,
recombination) is given by

\begin{equation}
G_{LO}^{E}=\frac{2}{\hbar }\sum_{{\bf n},occ}\gamma 
\mathop{\rm Im}%
\left\{ p_{j}^{{\bf n^{+*}}}p_{j}^{{\bf n^{-}}}\right\} \sqrt{N},
\label{eG_LO}
\end{equation}

where $N\equiv \left( n_{1}+1\right) \left( n_{2}+1\right) \left(
n_{3}+1\right) $.

The decay rate of the $\widetilde{LO}$ population turns out to be

\begin{equation}
D_{\widetilde{LO}}=2\kappa _{2}\sum_{{\bf n},j}\left\{ p_{j}^{{\bf n^{*}}%
}p_{j}^{{\bf n}}\right\} n_{2}=2\kappa _{2}\overline{n_{2}}.  \label{D_LO}
\end{equation}

Finally the escape rate of the $TA$ phonons is given by

\begin{equation}
E_{TA}=2\kappa _{3}\sum_{{\bf n},j}\left\{ p_{j}^{{\bf n^{*}}}p_{j}^{{\bf n}%
}\right\} n_{3}=2\kappa _{3}\overline{n_{3}}.  \label{E_TA}
\end{equation}

A diagram for the input and output electron rates is shown in figure \ref
{Diag2-Fig4}.

The kinetic equations (\ref{kin_eq}) can be approximated by a system
involving only populations. To do that we have to replace the amplitudes $%
p_{j}^{{\bf n}}$ from (\ref{dPj}) and (\ref{dP}) (See the appendix for a
detailed calculation). Then, we get a new set of kinetic equations

\begin{mathletters}
\label{kin_equations}
\begin{eqnarray}
&&{{\frac{d\overline{n_{1^{\prime }}}}{dt}}}=G-R_{1^{\prime }}\overline{%
n_{1^{\prime }}}-R_{w}  \label{Nm0} \\
&&{{\frac{d\overline{n_{0}}}{dt}}}=-R_{0}\overline{n_{0}}+R_{w}  \label{N0}
\\
&&{{\frac{d\overline{n_{1}}}{dt}}}=R_{w}-R_{\gamma }  \label{N1} \\
&&{{\frac{d\overline{n_{2}}}{dt}}}=-\kappa _{2}\overline{n_{2}}+R_{\gamma }
\label{N2} \\
&&{{\frac{d\overline{n_{3}}}{dt}}}=-\kappa _{3}\overline{n_{3}}+R_{\gamma }
\label{N3}
\end{eqnarray}

where

\end{mathletters}
\begin{eqnarray}
R_{w} &=&w\left[ \overline{n_{1^{\prime }}}\left( \overline{n_{1}}+1\right) -%
\overline{n_{0}}\ \overline{n_{1}}\right]  \label{w} \\
R_{\gamma } &=&\gamma _{o}[\overline{n_{1}}\left( \overline{n_{2}}+1\right)
\left( \overline{n_{3}}+1\right) -\left( \overline{n_{1}}+1\right) \overline{%
n_{2}}\ \overline{n_{3}}].  \label{gamma0}
\end{eqnarray}

$G$ is the input current, $R_{1^{\prime }}$ and $R_{0}$ are the escape rates
for one electron through the right barrier and $w$ is the transition rate
due to $LO$ phonon emission estimated in the appendix (and more roughly in 
\cite{jpcm}). The other parameters are: the decay rate of a $LO$ phonon $%
\gamma _{o}$, the decay rate of a $\widetilde{LO}$ phonon $\kappa _{2}$ and
the $TA$ phonon escape rate $\kappa _{3}$.

It can be seen in the expressions (\ref{w}) and (\ref{gamma0}) that the Bose
factors were obtained. That means that the phonons are produced in a
stimulated regime. For the case of $TA$ phonons, the Bose factor $\overline{%
n_{3}}+1$ is thousands times greater than the others.

\section{Results and conclusions}

The results presented here correspond only to the stationary case, i.e., $%
p_{j}^{{\bf n}}\left( t\right) =a_{j}^{{\bf n}}\left( \omega \right)
e^{i\omega t}$. In this case, all the populations (\ref{pop1}) and (\ref
{pop2}), and the rates (\ref{iG0`})-(\ref{E_TA}) are time independent and they
can be written as the same expressions of $a_{j}^{{\bf n}}\left( \omega
\right) $ instead of  $p_{j}^{{\bf n}}\left( t\right) $. The amplitudes 
$a_{j}^{{\bf n}}\left( \omega \right)$ are the solutions of the system (\ref{Amplitudes}).

The following parameters were used in our calculations. The barriers are $%
42.4\ ${\it \AA } wide and $300\ meV$ high ($38\ \%$ of {\it Al}
concentration). The well has a width of $203.4\ ${\it \AA }. The Fermi
energy at the emitter was fixed at $15\ meV$. The values estimated for the
decay rate $\kappa _{2}$ and the escape rate $\kappa _{3}$ were $30\ ps^{-1}$
and $0.05\ ps^{-1}$ respectively. The other parameters used are: $w$, that
depends on the applied bias but it is of the order of $2.5\ ps^{-1}$, $%
\gamma _{o}=0.11\ ps^{-1}$ taken from \cite{Vall}, and $g$, that also
depends on the applied bias and it is of the order of $3.2\ ps^{-1}$. The
value of $\gamma $ was taken equal to $0.65\ meV$ and was calculated from
the experimental value of $\gamma _{o}$ using \cite{Klemens} and \cite{Ziman}%
.

The transmittance through the DBH is calculated from

\begin{equation}
T={{\frac{\left| A_{T}^{0}\right| ^{2}+\left| A_{T}^{+}\right| ^{2}+\left|
A_{T}^{-}\right| ^{2}}{\left| A_{I}\right| ^{2}}}}.  \label{T}
\end{equation}

A plot of the transmittance as a function of energy is shown in figure \ref
{Transmit-Fig5}. It can be seen that when the ph-ph interaction is
considered (taking $\gamma \neq 0$), a central peak appears. The other two
peaks correspond to the polaronic branches already shown in previous works 
\cite{Tuyarot}, \cite{Weber}.

Once the transmission probabilities through each channel are obtained, the
output electronic current (equal to the input current at the stationary
regime) through the structure can be calculated by summing over the occupied
states at the emitter. This sum can be easily transformed into the integral

\begin{equation}
G\left( V\right) ={{\frac{S}{2\pi ^{2}}}}{{\frac{m^{\ast }}{\hbar ^{3}}}}%
\int\limits_{0}^{\varepsilon _{F}^{L}}{{\frac{\left( \varepsilon
_{F}^{L}-\varepsilon \right) j\left( \varepsilon \right) \rho \left(
\varepsilon \right) }{1+R+T}}}d\varepsilon  \label{G}
\end{equation}

where $S=0.5\cdot 10^{-3}\,mm^{2}$ is the device area, $m^{\ast
}=0.067\,m_{e}$ is the effective electron mass, $R$ is the reflectance and

\begin{eqnarray*}
j\left( \varepsilon \right) &=&{{\frac{4v}{\hbar }}}{{\frac{1}{\left|
A_{I}\right| ^{2}}}}\bigg\{\left| A_{T}^{0}\right| ^{2}\sin \left(
k_{0}^{^{\prime }}a\right) +\left| A_{T}^{+}\right| ^{2}\sin \left(
k^{+}a\right) +\left| A_{T}^{-}\right| ^{2}\sin \left( k^{-}a\right) \bigg\},
\\
\rho \left( E\right) &=&{{\frac{1}{2v\sin \left( k_{0}a\right) }}}.
\end{eqnarray*}

Instead of using the expression (\ref{G}) to calculate the electronic
current $G$, it is calculated from the sum of the terms (\ref{iG0`}) and (%
\ref{iG0}).

The difference between the currents calculated for $\gamma \neq 0$ and $%
\gamma =0$ is appreciable only in the dashed region of figure \ref
{current-Fig6}, that it is amplified in the inset.

Finally, the saser intensity and the $TA$ phonon population are shown in
figure \ref{saser-Fig7}. This curve is obtained by solving the kinetic
equations (\ref{kin_equations}).

In conclusion it was obtained here a set of kinetic equations using a full
quantum treatment. This set is different from those presented in a previous
work \cite{jpcm} where the kinetic equations were derived
phenomenologically. Besides, we consider for the first time all the relevant
phonons and its interactions in the calculation of the current.

It was shown that when the phonon-phonon interaction is considered, only
small modifications are observed for the current.

The accurate full quantum calculations for a DBH phonon laser confirm the
results obtained using rougher phenomenological methods \cite{jpcm}.

\section*{acknowledgments}

One of us (IC) wants to acknowledge Latin American agency CLAF for financial
support.

\appendix
\section*{}

In this appendix, we will show the way to obtain each term of equations (\ref
{kin_equations}) from (\ref{kin_eq}). To illustrate the procedure, we will
limit us to show only two examples.

We begin with the term

\begin{equation}
G_{1^{\prime }}^{O}=\frac{2}{\hbar }\sum_{{\bf n},occ}v_{1^{\prime }1}%
\mathop{\rm Im}%
\left\{ p_{1}^{{\bf n^{*}}}p_{1^{\prime }}^{{\bf n}}\right\} \text{.}
\label{Appendix-01}
\end{equation}

The channel $000$ is the only one that contribute significantly to the sum
appearing in (\ref{Appendix-01}) (see Fig. \ref{diagram_He} and Fig. \ref
{Diag1-Fig3}). Taking this into account, the expression (\ref{Appendix-01})
can be rewriten as

\begin{equation}
G_{1^{\prime }}^{O}=\frac{2}{\hbar }\sum_{occ}v_{1^{\prime }1}%
\mathop{\rm Im}%
\left\{ p_{1}^{0{\bf ^{*}}}p_{1^{\prime }}^{0}\right\} \text{.}
\label{Appendix-1}
\end{equation}

As the amplitudes appearing in (\ref{Appendix-1}) are time dependent, we
substitute one of them by their Fourier transform using (\ref{Fourier1})

\begin{equation}
\mathop{\rm Im}%
\left\{ p_{1}^{0{\bf ^{*}}}p_{1^{\prime }}^{0}\right\} =%
\mathop{\rm Im}%
\left\{ {{\frac{1}{\sqrt{2\pi }}}}\int_{-\infty }^{\infty }d\omega
a_{1}^{0^{*}}\left( \omega \right) e^{i\omega t}p_{1^{\prime }}^{0}\right\} .
\label{Appendix-2}
\end{equation}

To obtain an expression for the amplitude $a_{1}^{0^{\ast }}\left( \omega
\right) $, we use (\ref{A1_0}) and neglect the term containing the
renormalized hopping $\widetilde{v}_{\overline{1}1}^{0}$ in it. Taken into
account the dispersion relation (\ref{Disp_k000}), (\ref{e3}) and using (\ref
{At000}) to calculate the amplitude in the plane $j=2$, we get

\begin{equation}
a_{1}^{0^{\ast }}\left( \omega \right) =\Xi _{1^{\prime }1}a_{1^{\prime
}}^{0^{\ast }}\left( \omega \right)  \label{Appendix-3}
\end{equation}
where

\[
\Xi _{1^{\prime }1}=v_{1^{\prime }1}{{\frac{\left( \Sigma
_{1}^{0}-ve^{-ik_{0}^{^{\prime }}a}\right) }{\left[ \Sigma
_{1}^{0^{2}}-2v\Sigma _{1}^{0}\cos \left( k_{0}^{^{\prime }}a\right)
+v^{2}\right] }.}} 
\]

Replacing (\ref{Appendix-3}) in (\ref{Appendix-2})

\begin{equation}
\mathop{\rm Im}%
\left\{ p_{1}^{0{\bf ^{*}}}p_{1^{\prime }}^{0}\right\} ={{\frac{1}{\sqrt{%
2\pi }}}}%
\mathop{\rm Im}%
\bigg\{\int_{-\infty }^{\infty }\Xi _{1^{\prime }1}a_{1^{\prime
}}^{0^{*}}\left( \omega \right) d\omega e^{i\omega t}p_{1^{\prime }}^{0}%
\bigg\}.  \label{Appendix-4}
\end{equation}

The term $\Xi _{1^{\prime }1}$ inside the integral varies very slowly in the
range where the amplitude $a_{1^{\prime }}^{0^{\ast }}\left( \omega \right) $
is not null. Therefore, we can take it out of the integral. Now, performing
the integral over $\omega $ and taking the imaginary part we get

\begin{equation}
G_{1^{\prime }}^{O}=\frac{2}{\hbar }\sum_{occ}v_{1^{\prime }1}^{2}{{\frac{v}{%
\left[ \Sigma _{1}^{0^{2}}-2v\Sigma _{1}^{0}\cos \left( k_{0}^{^{\prime
}}a\right) +v^{2}\right] }}}\sin \left( k_{0}^{^{\prime }}a\right) \left|
p_{1^{\prime }}^{0}\right| ^{2}.  \label{Appendix-5}
\end{equation}

Defining $\overline{n_{1^{\prime }}}\equiv \sum_{occ}\left| p_{1^{\prime
}}^{0}\right| ^{2}$ we get

\[
G_{1^{\prime }}^{O}=R_{1^{\prime }}\overline{n_{1^{\prime }}} 
\]
where

\[
R_{1^{\prime }}={{\frac{2}{\hbar }}}v_{1^{\prime }1}^{2}{{\frac{v}{\left[
\Sigma _{1}^{0^{2}}-2v\Sigma _{1}^{0}\cos \left( k_{0}^{^{\prime }}a\right)
+v^{2}\right] }}}\sin \left( k_{0}^{^{\prime }}a\right) . 
\]

When the renormalization efects are not taken into account, i.e., $\Sigma
_{1}^{0}=0$, the expression obtained here for $R_{1^{\prime }}$ is the same
as that appearing in \cite{Sols1}.

The other term to be analyzed here is the rate of $LO$ phonon emission (\ref
{eG0`})

\begin{equation}
G_{1^{\prime }}^{E}=\frac{2}{\hbar }\sum_{{\bf n},occ}g%
\mathop{\rm Im}%
\left\{ p_{0}^{{\bf n}^{+*}}p_{1^{\prime }}^{{\bf n}}\right\} \sqrt{n_{1}+1}.
\label{Appendix-7}
\end{equation}

First, we expand the term inside the sum and change the range over which the 
$n_{1}$ runs to obtain (\ref{w}). That is

\begin{equation}
\sum_{{\bf n}}%
\mathop{\rm Im}%
\left\{ p_{0}^{{\bf n}^{+*}}p_{1^{\prime }}^{{\bf n}}\right\} \sqrt{n_{1}+1}%
=\sum_{{\bf n}}i\Big(p_{0}^{\left( n_{1}+1\right)
n_{2}n_{3}^{*}}p_{1^{\prime }}^{{\bf n}}\sqrt{n_{1}+1}-p_{1^{\prime
}}^{\left( n_{1}-1\right) n_{2}n_{3}^{*}}p_{0}^{{\bf n}}\sqrt{n_{1}}\Big).
\label{Appendix-8}
\end{equation}

Writing $p_{0}^{\left( n_{1}+1\right) n_{2}n_{3}^{*}}$ in the first term of (%
\ref{Appendix-8}) in terms of its Fourier transform $a_{0}^{\left(
n_{1}+1\right) n_{2}n_{3}^{*}}$, doing the same with $p_{1^{\prime
}}^{\left( n_{1}-1\right) n_{2}n_{3}^{*}}$ in the second term and using
equations (\ref{1A0}), (\ref{1Am0}) (only the terms that contain the factor $%
g$) we get

\begin{equation}
a_{0}^{\left( n_{1}+1\right) n_{2}n_{3}}={{\frac{g}{\Delta _{0}^{\left(
n_{1}+1\right) n_{2}n_{3}}}}}\sqrt{n_{1}+1}a_{1^{\prime }}^{{\bf n}}
\label{Appendix-9}
\end{equation}

and

\begin{equation}
a_{1^{\prime }}^{\left( n_{1}-1\right) n_{2}n_{3}}={{\frac{g}{\Delta
_{1^{\prime }}^{\left( n_{1}-1\right) n_{2}n_{3}}}}}\sqrt{n_{1}}a_{0}^{{\bf n%
}},  \label{Appendix-10}
\end{equation}

where

\[
\Delta _{0}^{\left( n_{1}+1\right) n_{2}n_{3}}=\hbar \omega -\left(
n_{1}+1\right) \hbar \omega _{1}-n_{2}\hbar \omega _{2}-n_{3}\hbar \omega
_{3}-\varepsilon _{0} 
\]

and

\[
\Delta _{1^{\prime }}^{\left( n_{1}-1\right) n_{2}n_{3}}=\hbar \omega
-\left( n_{1}-1\right) \hbar \omega _{1}-n_{2}\hbar \omega _{2}-n_{3}\hbar
\omega _{3}-\varepsilon _{1^{\prime }}. 
\]

Replacing (\ref{Appendix-9}) and (\ref{Appendix-10}) in (\ref{Appendix-7})
and doing the inverse Fourier transform for the amplitudes with the same
approximation as those made to get equation (\ref{Appendix-5})

\begin{equation}
G_{1^{\prime }}^{E}=\sum_{{\bf n},occ}\Big\{{{\frac{g^{2}}{\hbar \Delta
_{0}^{\left( n_{1}+1\right) n_{2}n_{3}}}}}\left( n_{1}+1\right) \left|
p_{1^{\prime }}^{{\bf n}}\right| ^{2}-{{\frac{g^{2}}{\hbar \Delta
_{1^{\prime }}^{\left( n_{1}-1\right) n_{2}n_{3}}}}}n_{1}\left| p_{0}^{{\bf n%
}}\right| ^{2}\Big\}.  \label{Appendix-11}
\end{equation}

If we do now a mean field approximation to take the phonon terms out of the
sums, we get the following expression

\begin{equation}
G_{1^{\prime }}^{E}=\left( \overline{n_{1}}+1\right) \sum_{{\bf n},occ}{{%
\frac{g^{2}}{\hbar \Delta _{0}^{\left( n_{1}+1\right) n_{2}n_{3}}}}}\left|
p_{1^{\prime }}^{{\bf n}}\right| ^{2}-\overline{n_{1}}\sum_{{\bf n},occ}{{%
\frac{g^{2}}{\hbar \Delta _{1^{\prime }}^{\left( n_{1}-1\right) n_{2}n_{3}}}}%
}\left| p_{0}^{{\bf n}}\right| ^{2}.  \label{Appendix-12}
\end{equation}

The amplitudes $p_{1^{\prime }}^{000}$ and $p_{0}^{100}$ are the only ones
that contribute significantly to the first and second sum, respectively.
Taking this into account and the fact that close to the resonance ($%
\varepsilon _{1^{\prime }}-\varepsilon _{0}\simeq $ $\hbar \omega _{1}$) $%
\Delta _{1^{\prime }}^{0}\simeq \Delta _{0}^{100}=\Delta $, equation (\ref
{Appendix-12}) can be written as

\begin{equation}
G_{1^{\prime }}^{E}=\left( \overline{n_{1}}+1\right) \sum_{occ}\frac{g^{2}}{%
\hbar \Delta }\left| p_{1^{\prime }}^{0}\right| ^{2}-\overline{n_{1}}%
\sum_{occ}\frac{g^{2}}{\hbar \Delta }\left| p_{0}^{100}\right| ^{2}
\label{Appendix-13}
\end{equation}

Finally the expression (\ref{Appendix-13}) can be written as

\begin{eqnarray}
&&G_{1^{\prime }}^{E}\simeq w\left[ \overline{n_{1^{\prime }}}\left( 
\overline{n_{1}}+1\right) -\overline{n_{0}}\ \overline{n_{1}}\right] \equiv
R_{w}  \label{Appendix-14} \\
&&\text{where }w=\sum_{occ}\frac{g^{2}}{\hbar \Delta },  \nonumber \\
&&\overline{n_{0}}\equiv \sum_{occ}\left| p_{0}^{100}\right| ^{2}\text{ and }%
\overline{n_{1^{\prime }}}\equiv \sum_{occ}\left| p_{1^{\prime }}^{0}\right|
^{2}.  \nonumber
\end{eqnarray}

The other terms appearing in (\ref{kin_eq}) can be transformed by the same
procedure described above.

In summary, to obtain the kinetic equations similar to that used in \cite
{jpcm}, i.e., involving only populations, we did not take into account the
crossed terms. For example, if we consider the term with $v_{\overline{1}0}$
in (\ref{1A0}) to obtain (\ref{Appendix-9}), then the expression for the
amplitude $a_{0}^{\left( n_{1}+1\right) n_{2}n_{3}}$ would have a term
depending on $v_{\overline{1}0}$. That term would describe a virtual
process. In this process, an electron tunnel through the left barrier ($j=%
\overline{1}$) directly to the ground state ($j=0$) emitting one $LO$
phonon. Even when this process could occur, its probability is much smaller
than that of the indirect process: first tunnel to the excited level, and
then decay to the ground state emitting a $LO$ phonon. Therefore, we make
the sum of each process as being independent. The results obtained from
equations (\ref{kin_equations}) are quite similar to those obtained from (%
\ref{kin_eq}).

\begin{figure}[tbph]
\caption{Potential profile and energy levels at the resonant condition. $%
b_{l}(b_{r})$ and $d$ are the barriers and well width respectively.}
\label{profile-Fig1}
\end{figure}

\begin{figure}[tbph]
\caption{The profile for the scattering region.}
\label{well-Fig2}
\end{figure}

\begin{figure}[tbph]
\caption{Diagram representing the electronic Hamiltonian (11).}
\label{diagram_He}
\end{figure}

\begin{figure}[tbph]
\caption{Diagram representing equations (29). The vertical transition
represents the $LO$ phonon emission (absorption). As the channels $+$ and $-$
have the same energy, they are represented in the same plane. The gray
points represent the sites with the renormalized energies and the gray lines
are the renormalized hoppings.}
\label{Diag1-Fig3}
\end{figure}

\begin{figure}[tbph]
\caption{Diagram for the several rates appearing in (41a) and (41b).}
\label{Diag2-Fig4}
\end{figure}

\begin{figure}[tbph]
\caption{Dashed line: the two polaronic peaks when $\gamma =0$. Solid line:
if we assume $\gamma \neq 0$, it appears a third peak corresponding mainly
to one electron in the ground state with a pair ($\widetilde{LO}+TA$) of
phonons.}
\label{Transmit-Fig5}
\end{figure}

\begin{figure}[tbph]
\caption{Solid line: the current for $\gamma \neq 0.$ For comparison, the
electronic current when $\gamma =0$ is shown in the inset (dashed line). It
appears a shoulder in $G$ ($\gamma \neq 0$) corresponding to the third peak
of figure 6. The cross-sectional area used was $S=0.5\cdot 10^{-3}\;mm^{2}$.}
\label{current-Fig6}
\end{figure}

\begin{figure}[tbph]
\caption{The $TA$ phonon population $\overline{n_{3}}$ and the saser
intensity $S=\kappa _{3}\overline{n_{3}}$.}
\label{saser-Fig7}
\end{figure}

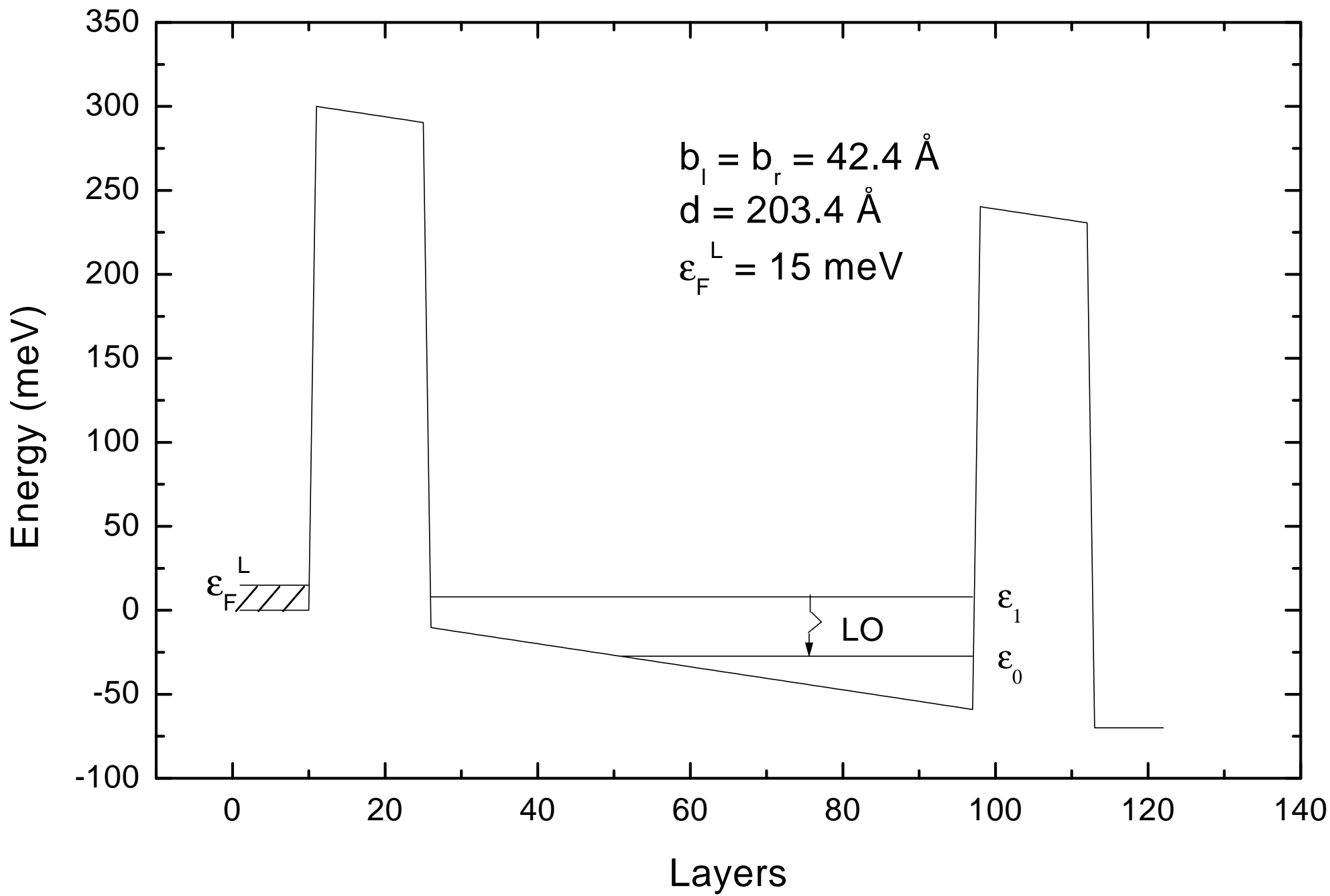
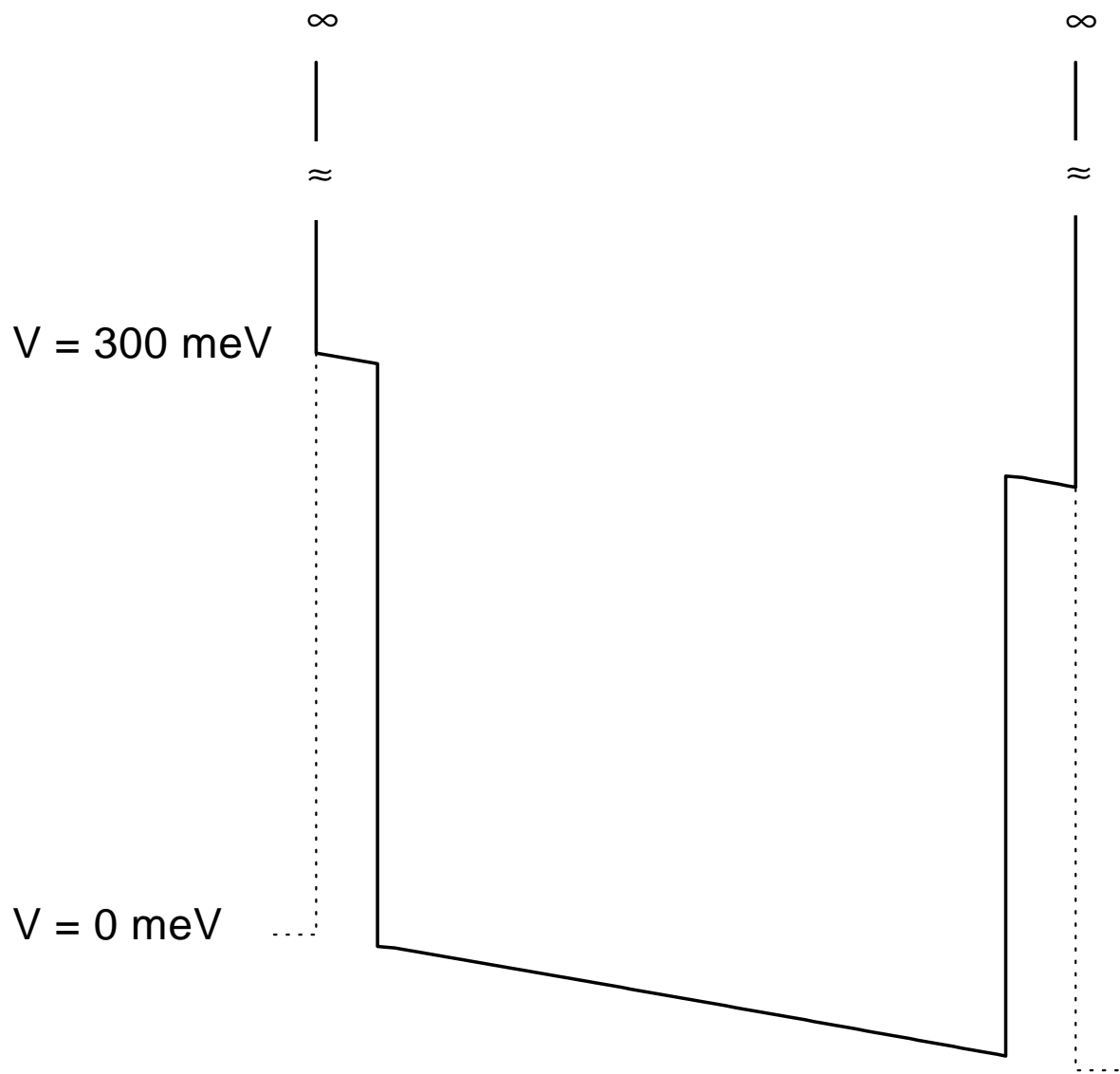
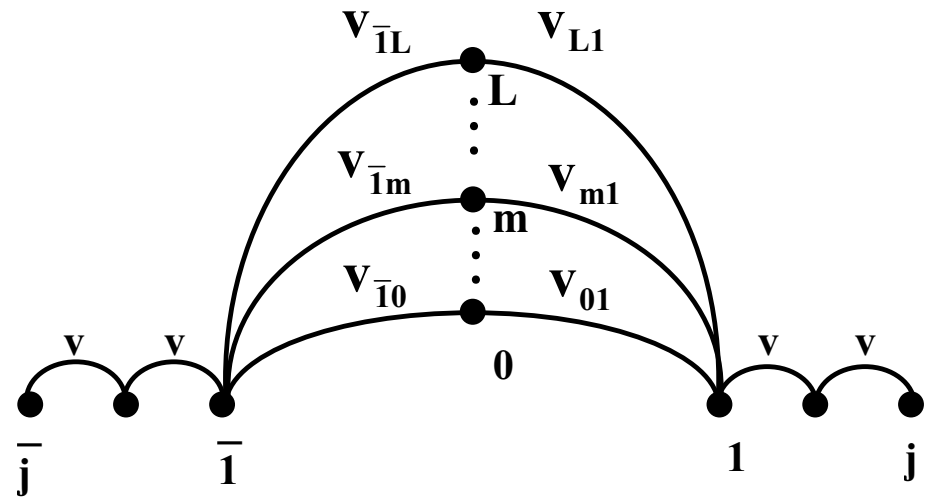
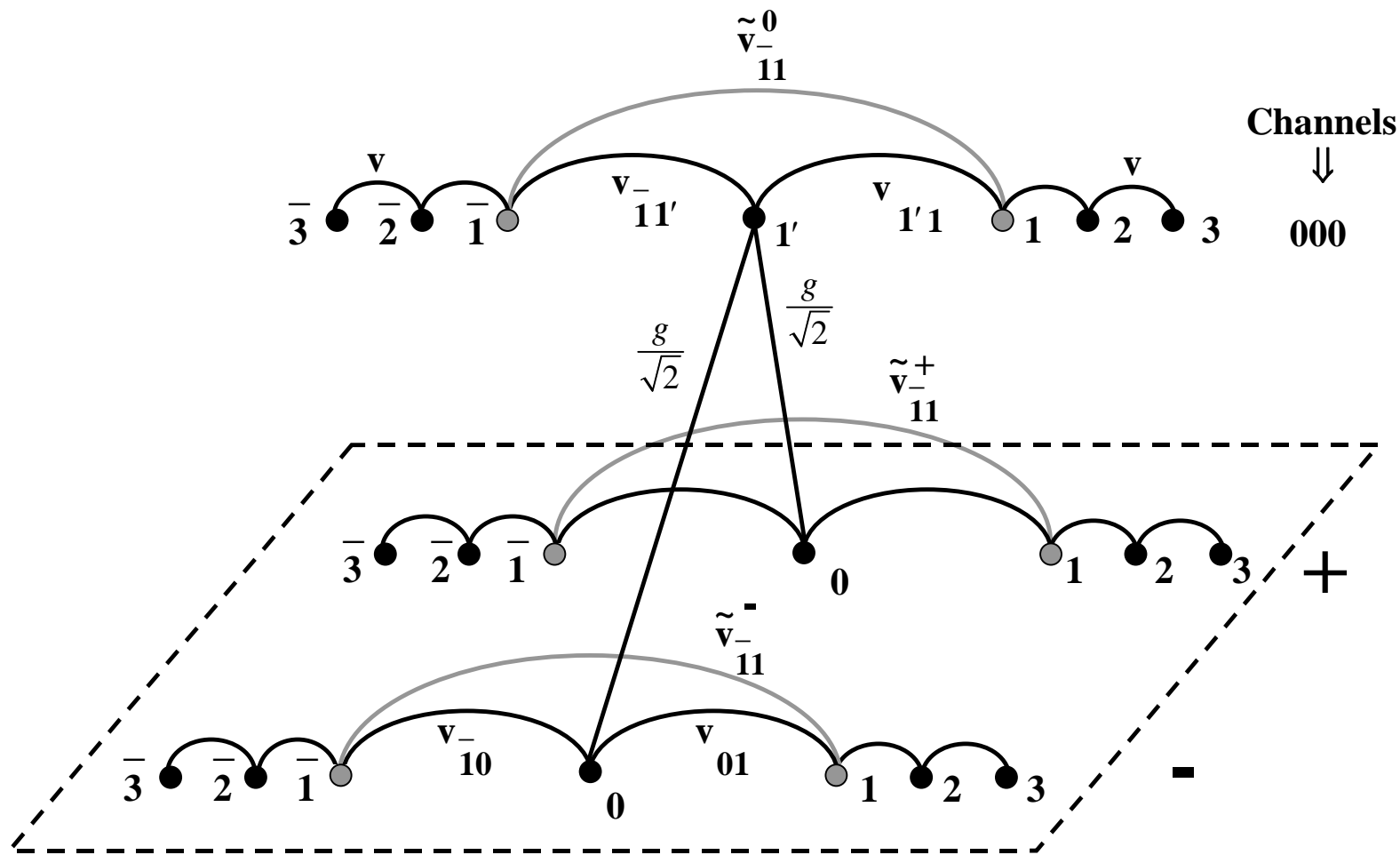
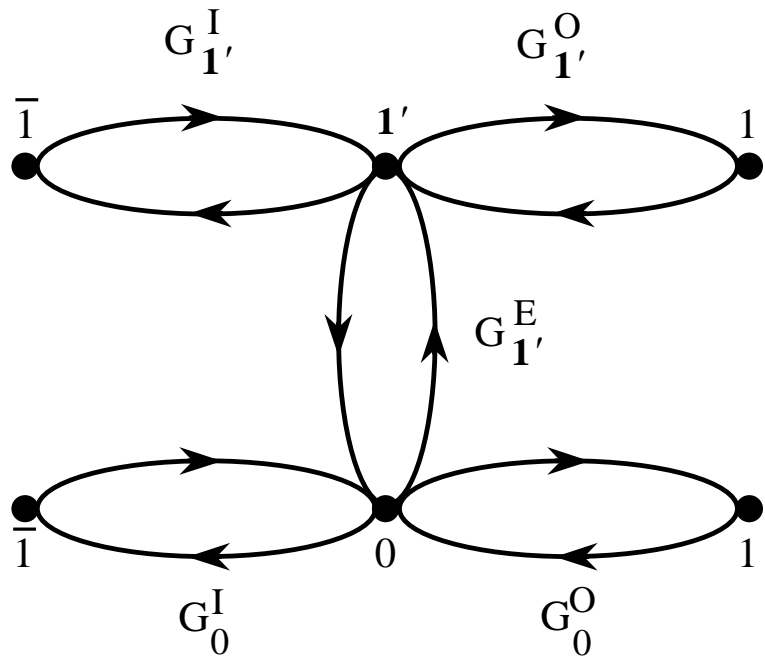
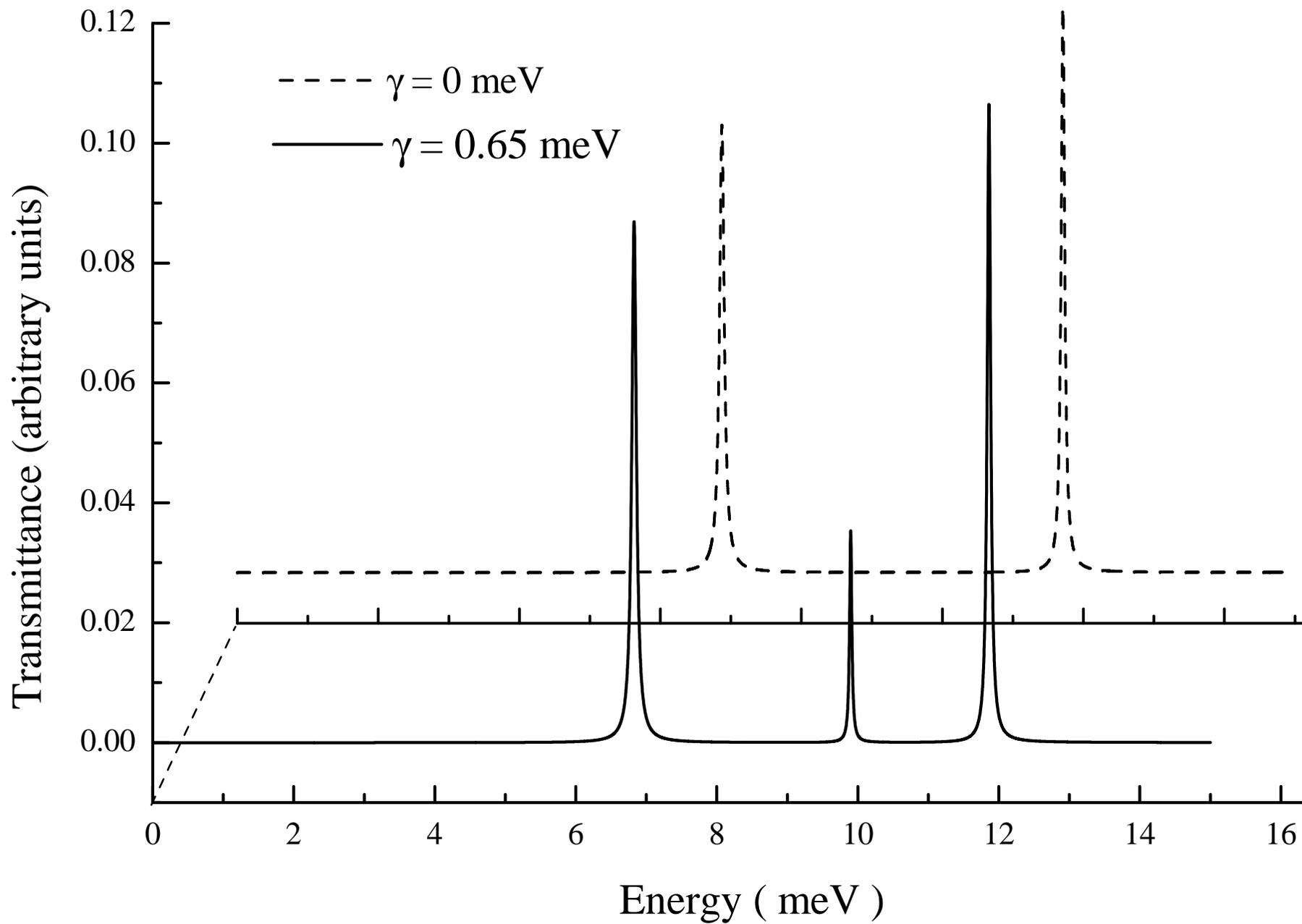
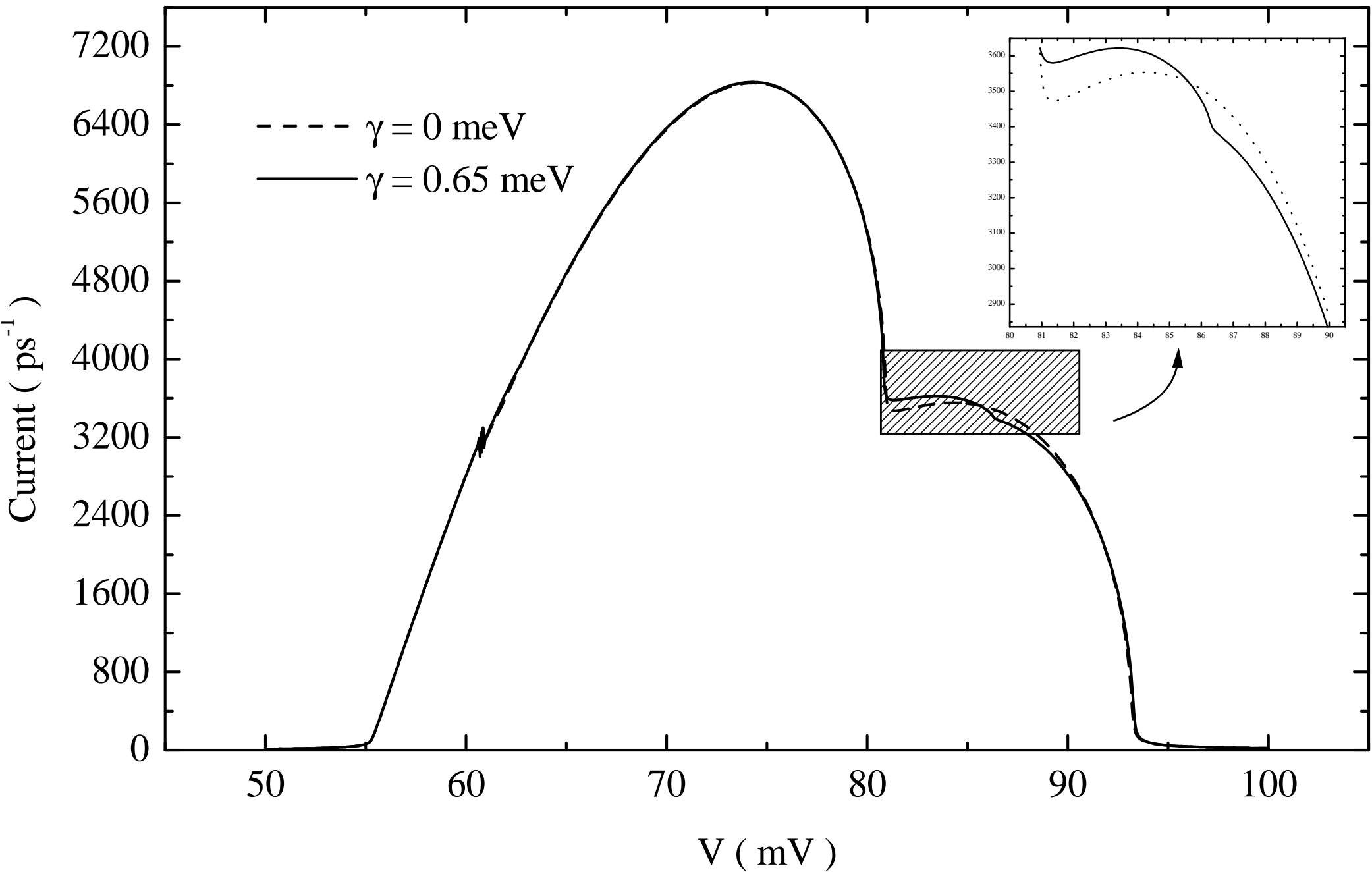
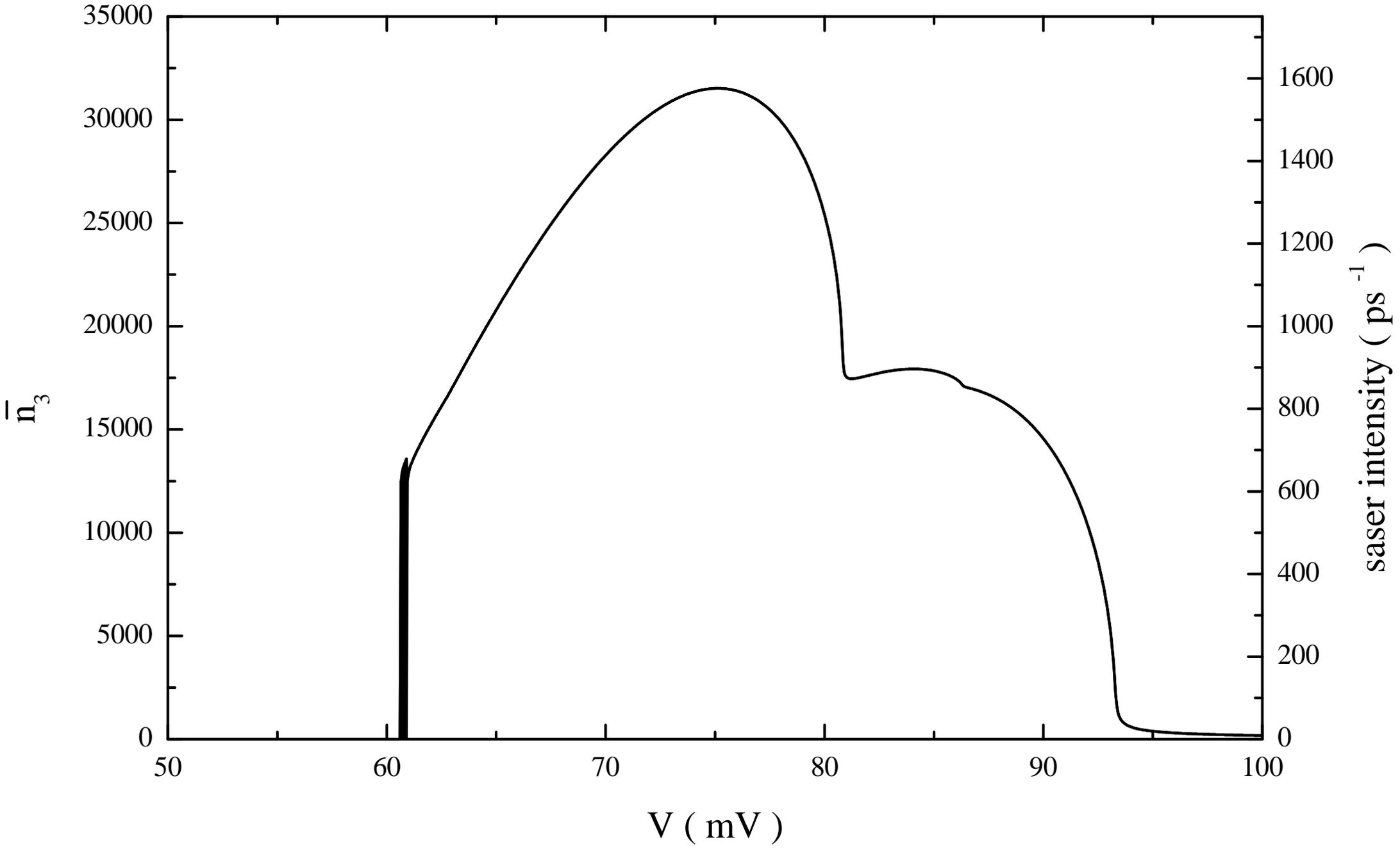
\end{document}